\documentclass{freiamemo}

\usepackage{graphicx}
\usepackage{dcolumn}
\usepackage{bm}
\usepackage{amssymb}
\usepackage{amsmath}
\usepackage{siunitx}
\usepackage{subcaption}
\usepackage{array}
\usepackage{rotating}
\usepackage{color}
\usepackage{tabularx}
\usepackage{hyperref}
\usepackage[english]{babel}
\usepackage[utf8x]{inputenc}
\usepackage[T1]{fontenc}
\usepackage[dvipsnames]{xcolor}
\expandkeywordstrue

\begin{document}

\title{First magnet operation on the cryogenic test stand Gersemi at FREIA}

\author{K\'evin Pepitone, Konrad Gajewski, Lars Hermansson, Roc\'io Santiago Kern}
\email{kevin.pepitone@physics.uu.se}
\departmentname{Physics and Astronomy}
\date{24-Aug-2021}
\documentlabel{FREIA Report 2021/01}
\keywords{Magnet test stand, Gersemi, FREIA laboratory}

\enablepageheader
\maketitle

\begin{abstract}
The Gersemi cryogenic test bench, installed at FREIA laboratory at Uppsala University, was used for the first time in 2021 to power a superconducting magnet. As part of the HL-LHC program, this cryostat offers an operating temperature between 4.2K and 1.9K. Its satellite equipment such as power converters and the acquisition system allow two superconducting magnet coils to be powered up to 2 kA and provide magnet protection through a robust quench detection and two energy extraction units. This report describes Gersemi's first cryogenic operational experiment and the safety strategy to ensure magnet integrity during operation.
\end{abstract}


\section{Introduction}

FREIA laboratory \cite{Ruber:2014iva}, in Uppsala University has various facilities \cite{ruber2021accelerator} including the vertical cryostat Gersemi \cite{Santiago-Kern1446639, thermeau:hal-02497837}. Initially commissioned with the liquid insert \cite{miyazaki2020cold}, the magnet insert was prepared and successfully used in 2020-2021 to perform the powering of a superconducting (SC) orbit corrector magnet. In this report we will first introduce, Sec.~\ref{Magnet insert}, the magnet insert, including the cabling and the cryogenic equipment. Sec.~\ref{Satellite equipment} will focus on the satellite equipment which are the power supplies, energy extraction units, quench detection system and acquisition system. Finally, Sec.~\ref{Commissioning and magnet test results} will focus on the commissioning at warm and at cold.

\section{Magnet insert} \label{Magnet insert}

\begin{figure}[tb]
\centering
\includegraphics[width = 16cm]{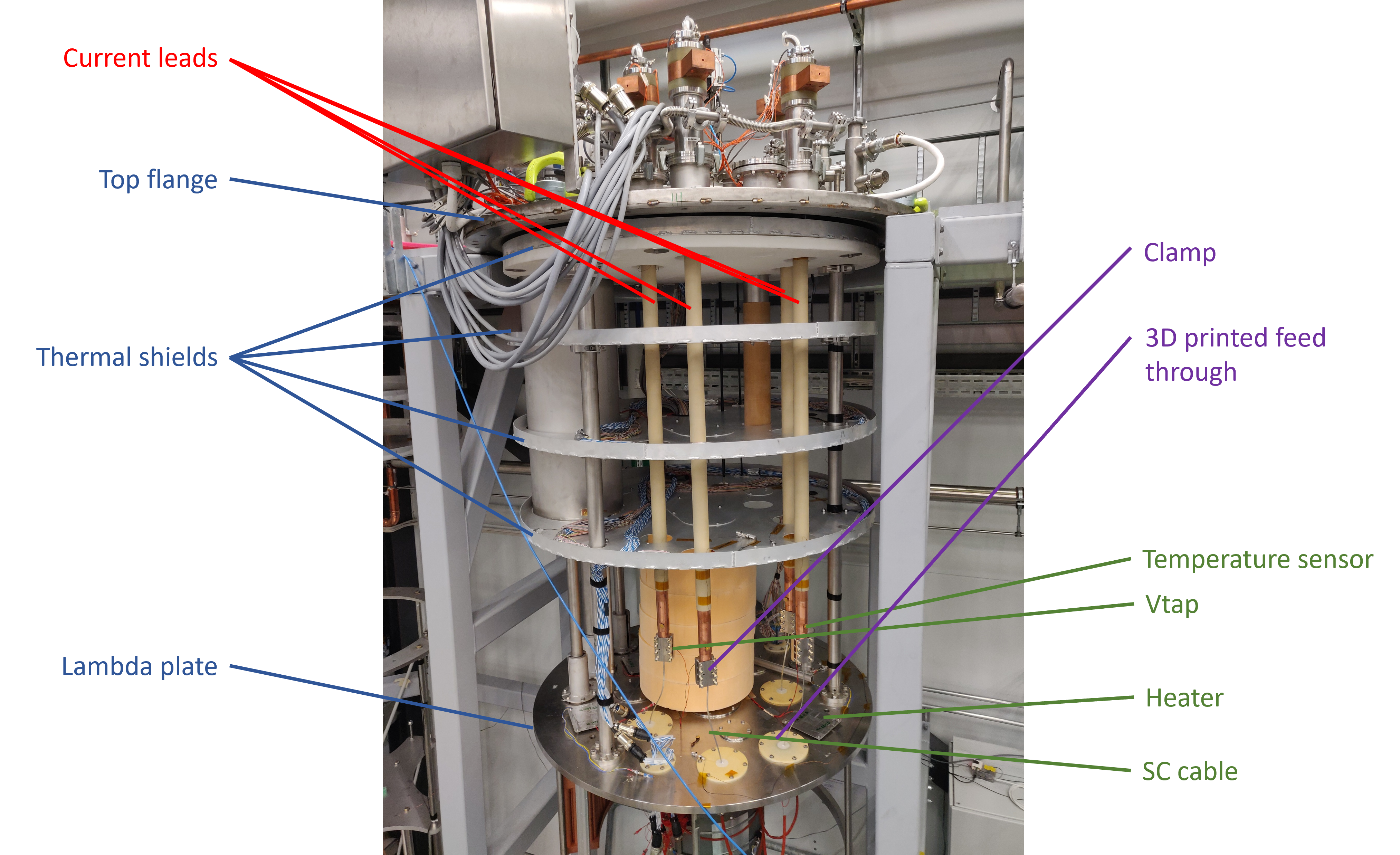}
\caption{Photograph of the magnet insert, above the lambda plate. Mechanical parts are shown in blue, current leads in red, an example of the cryogenic equipment in green and the extra parts in purple.}
\label{fig:Magnet_insert}
\end{figure}

The magnet insert offers the possibility of operating between 4.2 K and 1.9 K by maintaining the liquid helium bath at a pressure slightly above atmospheric \cite{Santiago-Kern1446639}. Fig. \ref{fig:Magnet_insert} shows an image of the magnet insert. The upper part is composed of cryogenic thermal shields on which the acquisition signals and instrumentation cables are resting. Specific silver plated copper cables are used, which are common in aeronautics or military equipment, since they offer a perfect resistance to cryogenic conditions. From the upper flange to the lambda plate, 4 current leads supporting up to 2 kA\textsubscript{DC} each are installed. Many cables, but also SC cables, attached to the current leads, pass through the lambda plate. While the latter must be leak tight, 3D printed and cryogenic resistant feedthroughs, visible on Fig. \ref{fig:Magnet_insert}, have been designed and adapted to the insert. A clamping system used as interface between current lead - SC cable above the lambda plate and between SC cable - SC cable below the lambda plate have been designed and successfully tested during the first operation of Gersemi. It allows to easily and safely connect the current lines from the magnets to the insert. A specific tool, not shown here, has been used to weld two layer of SC cable together, along the large surface, limiting the flexibility of the cable. A groove with a width equal to the one of the cable and a thickness slightly smaller was realized on the stainless steel clamps. The contact between SC cable - SC cables or SC - the current lead is made by applying pressure on the clamps and using indium wire to improve the conductivity.

Multiple cryogenic equipment, including a heat exchanger, thermometers, level probes, heaters, pressure gauges and valves are positioned on the magnet insert above and below the lambda plate \cite{Santiago-Kern1446639}.

Instrumentation cables to/from the magnet, commonly called voltage taps of Vtaps, are distributed above and below the lambda plate, and are used to monitor the voltage difference between two parts of the magnet windings or the insert. A total of eight of these Vtaps are placed above and below the lambda plate on the interface current leads - SC cables and SC cables - magnets and thirty-two in total can be used to monitor the magnet. These Vtaps are primarily used for quench detection, see Sec.~\ref{Quench detection system}.

\section{Satellite equipment} \label{Satellite equipment} 

\begin{figure}[tb]
\centering
\includegraphics[width = 16cm]{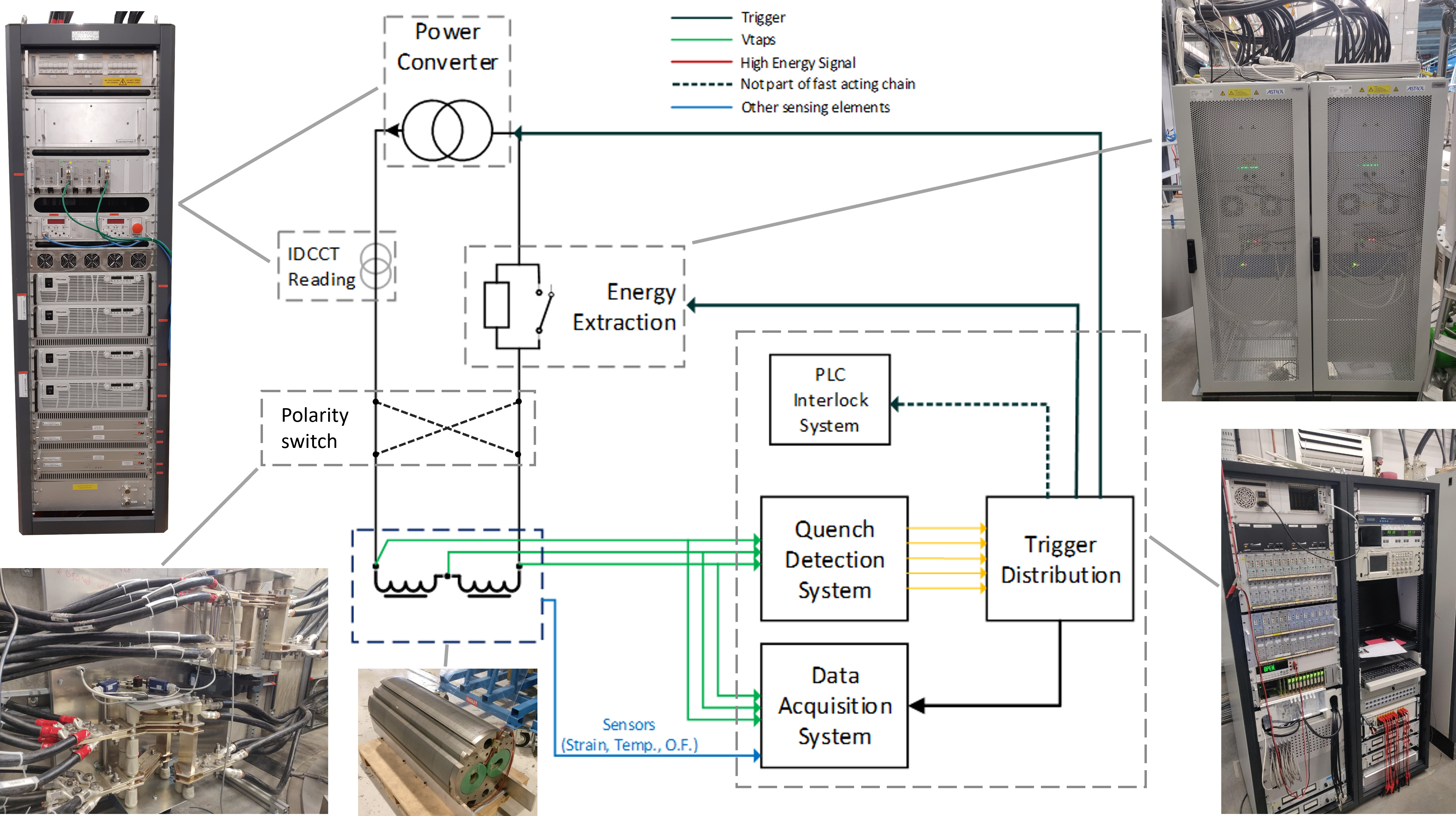}
\caption{Layout and pictures of the satellite equipment for the magnet tests.}
\label{fig:Layout_sat_eq}
\end{figure}

On a test bench such as Gersemi, a multitude of satellite equipment is used during the operation of SC magnets. We will now describe this equipment in four different subsections. The layout presented in Fig. \ref{fig:Layout_sat_eq} shows the satellite equipment and the connections between them and the magnet.

\subsection{Power supplies} \label{Power supplies}

The Fig. \ref{fig:Layout_sat_eq} top left presents the power supplies cabinet. This system also called COMBO (COMmercial Based cOnverter) \cite{COMBO} is part of a family of power supplies based on commercial units (off-the-shelf) only and has been developed at CERN. The main goal is to reproduce the same functionalities of a standard power supply, recreating if required some interfaces and function. The system installed at FREIA can provide a maximum of 2 x 2 kA\textsubscript{DC} for a maximum voltage of 10 V\textsubscript{DC} in one quadrant. In addition of this system a mechanical polarity reversing switch has been installed to operate in four quadrants. The power part of the COMBO power rack is completed by: (1) the CERN digital controller, so-called FGC3, which is a high level control system with high precision digital current loop. It collects and reports all status, faults, and measurements from all the different parts to the remote services, for diagnostic and operation purposes. (2) high precision DC Current Transformers (IDCCT), measuring DC or pulse current at the required precision.

\subsection{Energy extraction} \label{Energy extraction}

The energy extraction units (one for each 2 kA\textsubscript{DC} line) are shown in Fig. \ref{fig:Layout_sat_eq} top right. This water cooled system designed by CERN and manufactured by the industry is based on IGBTs (Insulated-Gate Bipolar Transistor) switches \cite{Dahlerup-Petersen:2622105}. The energy extraction units are based on two identical, series-connected IGBTs, for redundancy purposes, in a series-connection with a reverse-blocking power diode. The three heat sinks of the power semiconductors are the only elements which require water cooling. The IGBT’s are powered up to 1 kA\textsubscript{DC}, which is a significant reduction with respect to their normal rating (3600 A). With this choice no water cooling of the emitter / collector busbars are required. The complete units are built with a fully laminated busbar structure, reducing to a minimum the stray inductances and, herewith, the size of the compensating capacitance. Pluggable, spring-loaded fork connectors with multi-lam strips of CuBe louvers are used for the 1 kA power connections between the basic modules and the DC distribution bars at the back of the instrumentation rack.

Our energy extraction system can be associated with different dump resistors. The 4 units of each of the following resistors 300 m\si{\ohm}, 600 m\si{\ohm} and 700 m\si{\ohm} can provide a multitude of values between 77 m\si{\ohm} and 3200 m\si{\ohm} depending on the connections (series or parallel).

\subsection{Acquisition systems} \label{Acquisition systems}

\begin{figure}[tb]
\centering
\includegraphics[height = 12cm]{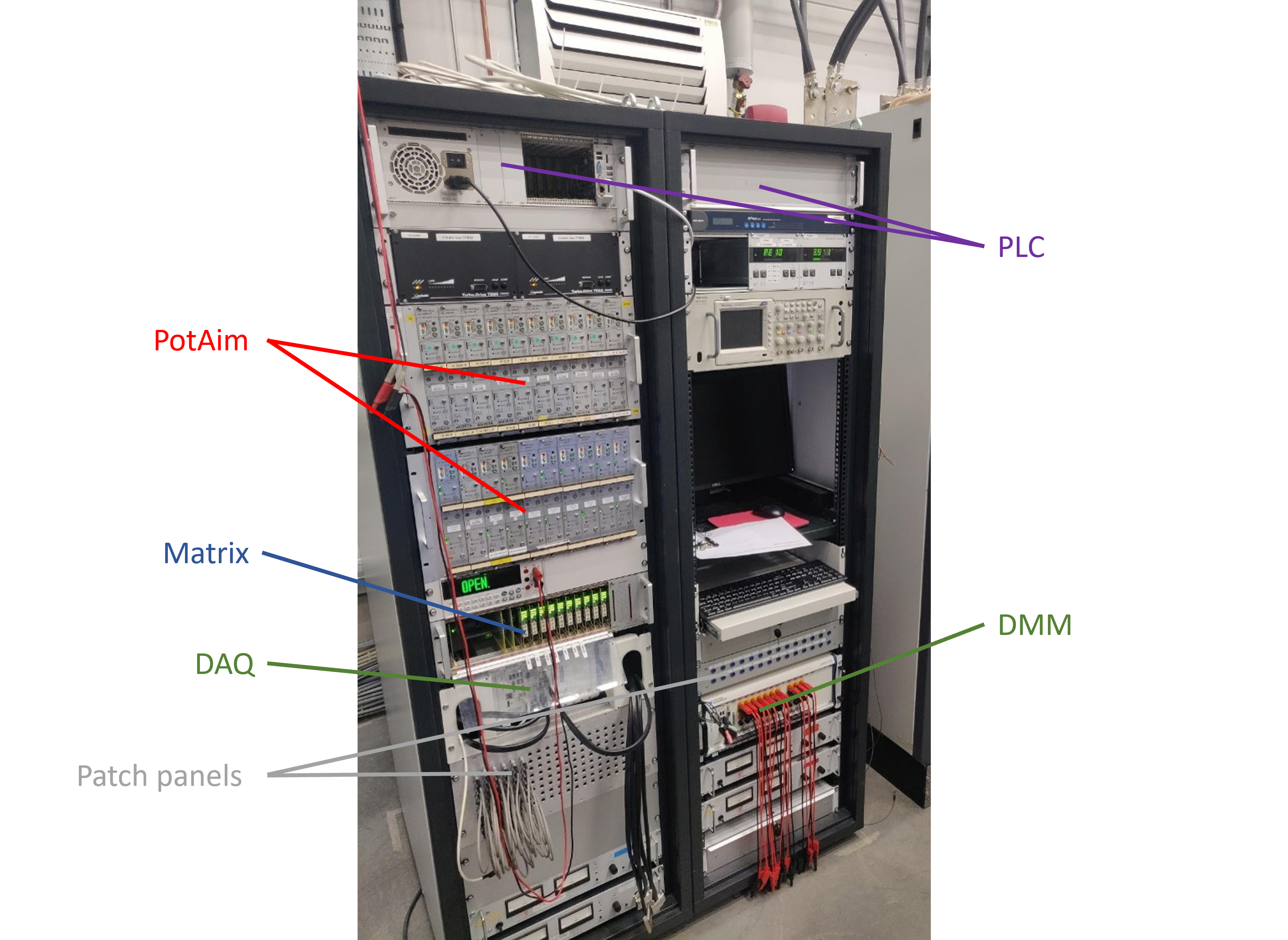}
\caption{DAQ, DMM and quench detection system. The acquisition system are shown in green, the safety parts in red, purple and blue, and the extra equipment in grey.}
\label{fig:DAQ_DMM_Quench}
\end{figure}

In Gersemi, two different acquisition systems are used called respectively DAQ and DMM (see Fig. \ref{fig:DAQ_DMM_Quench}). Both are based on the National Instruments system although the DAQ works like an oscilloscope, the DMM behaves like a voltmeter. 

The DAQ system is used to analyse the signals coming from the magnet during the training of the latter. When a quench is detected the DAQ is triggered, two type of cards are available (1) Low Frequency (LF) and (2) High Frequency (HF). The acquisition window is 1.2 s and 45 ms for the LF cards and HF cards respectively. We can monitor up to 72 signals with an acquisition frequency of 1 kHz (LF cards), 2 MS/s, 16-bit resolution and 64 signals with an acquisition frequency in the range 1-200 kHz (HF cards), 3.33 MS/s, 16-bit resolution with the DAQ system. 

In parallel to this system, the DMM system is used for live measurements such as RRR (Residual Resistivity Ratio), splices or inductance measurements. It offers the possibility to acquire up to 10 signals with a voltage measurement from ±10 nV to 1000 V\textsubscript{DC} and an accuracy of 7½-digit. The DMM software is such that acquisitions are not limited in time.

\subsection{Quench detection system} \label{Quench detection system}

The quench detection system is the most important satellite equipment and consists of Vtaps installed on the magnets or the insert, a system called matrix, another called PotAim (Potentiel Aimant) cards, and a safety PLC (see Fig. \ref{fig:DAQ_DMM_Quench}). We will now present these different systems.
\begin{itemize}
\item The Vtaps are wires installed at different places on the magnet and on the insert. By measuring the voltage difference between two Vtaps or a differential voltage between three Vtaps, it is possible to confirm that the magnet is quenching. A SC magnet has zero resistivity when it is in a superconducting state. If the differential voltage is not equal to 0 V, it means that a part of the magnet is resistive so it is quenching.
\item The PotAim cards are directly connected to the Vtaps, a voltage threshold and a validation time are fixed on them. If the voltage between two Vtaps (or the differential voltage between three Vtaps) is higher than a certain threshold for a time longer than a certain validation time, the quench is considered as validated, which causes a trigger signal to be sent by the PotAim card(s) to the matrix system.
\item The matrix system distributes the trigger signal to all the equipment: power supplies, energy extraction, DAQ, cryogenics and safety PLC in order to stop the power supply, to extract the energy stored in the magnet outside of the liquid helium bath (instead of inside) and to trigger the DAQ to then post-process the results.
\item The safety PLC is the "brain" of the quench detection system. It ensures that all conditions from the satellite equipment mentioned above and also from the cryogenics are correct to allow powering of the magnet. During the powering, it ensures that the satellite equipment is in good condition, otherwise it stops everything.
\end{itemize}

\section{Commissioning and magnet test results} \label{Commissioning and magnet test results}

The commissioning was divided into two parts, first we commissioned the high current line, consisting of the power converters, polarity reversing switches, energy extraction units and 32 copper cables each with a cross section of 150 mm$^2$. Then we commissioned at cold with a magnet.

We began testing the equipment on a short circuit at warm and produced some extraction at various currents from 0 to 2 kA. After successful operation, we measured the temperature rise in the copper cables by powering, again on the short, at 2 kA for 4 hours. The maximum temperature measured on the copper cable was below 60 \si{\celsius}. Our control system and safety PLC were also tested during this commissioning phase. The PLC showed the expected behavior and has been integrated into FREIA's control system (EPICS) allowing use of standard program for operator's interface and access to general services like archiving and alarms.

For the cold commissioning with the magnet insert, we used a LHC orbit corrector, so-called MCBC \cite{Bruning:782076}. Characteristics of this magnet are given in the following table \ref{tab:table1}.

\begin{table}
\caption{MCBC parameters.}
\begin{tabular}{ll}
\textbf{Parameters} & \textbf{Values} \\
\hline
Coil inner diameter & 56.4 mm \\
Magnetic length & 0.904 m \\
Nominal field (at 1.9 K / 4.2 K) & 3.11 T / 2.33 T \\
Nominal current (at 1.9 K / 4.2 K) & 100 A / 80 A \\
Short sample current (at 1.9 K / 4.2 K) & 172 A / 127 A \\
Stored energy (at 1.9 K) & 14.2 kJ \\
Self-inductance & 2.84 H \\
DC resistance (RT) & 375 \si{\ohm} \\
Ribbon construction & 14 wire glued \\
SC wire & NbTi in Cu matrix, 0.38 mm x 0.73 mm\\
\end{tabular}
\label{tab:table1}
\end{table}

After a cooling down period, during which we have been testing different sequences and the cryogenic control system, we reached the required temperature of 1.9 K to operate the magnet. We powered the magnet to a current equal to 80 A at 1.9 K and to 64 A at 4.2 K. These values correspond to 80 \% of the nominal current (limit requested by the magnet owner).
Ramp-ups and ramp-downs at 0.1 A/s were applied to the magnet to measure the inductance. We experimentally found an inductance of 2.8 H matching the specifications. The resistance of the entire system was measured at 24 m\si{\ohm}.

During cold commissioning, all the satellite equipment showed the expected behavior. The power converters in series with the polarity reversing switches and the power extraction units responded as expected. The safety matrix and PotAim cards helped us to protect the magnet and some forced energy extractions were performed.
All the connections made on the magnet insert above and below the lambda plate, both for the power lines with SC cables, the signals on the magnet and the probes needed for cryogenics worked perfectly as well as the acquisition systems. After warming up the cryostat and inspection of the insert, no significant damage was observed.

\section{Conclusion} \label{Conclusion}

The Gersemi vertical cryostat was successfully used in 2021 to operate a superconducting magnet. All equipment, acquisition systems, power converters, energy extraction units, polarity reversing switches, PotAim, matrix, insert, control system, and safety PLC performed as expected. No significant limitations were observed during the first commissioning at warm and at cold. Upgrades to the insert and cryostat are under way and will be tested soon. We are confident in using Gersemi for the HL-LHC program and for third party labs through ARIES. 

\bibliography{Gersemi.bib}
\bibliographystyle{unsrt}

\end{document}